\RequirePackage{fix-cm}
\documentclass[smallextended]{svjour3}       
\smartqed  
\usepackage{graphicx}
\begin{document}

\title{Thermal coupling of silicon oscillators in cryogen-free dilution refrigerators
}


\author{David Schmoranzer         \and
        Sumit Kumar \and
        Annina Luck \and
        Eddy Collin \and
        Andrew Fefferman \and
        Xiao Liu \and
        Thomas Metcalf \and
        Glenn Jernigan
}


\institute{David Schmoranzer \and Sumit Kumar \and Annina Luck \and Eddy Collin \and Andrew Fefferman \at
              Universit\'{e} Grenoble Alpes and Institut N\'{e}el, CNRS, Grenoble, France \\
              \email{andrew.fefferman@neel.cnrs.fr}           
           \and
           X. Liu \and T. Metcalf \and G. Jernigan \at
              Naval Research Laboratory, Washington, D.C., USA
}

\date{Received: date / Accepted: date}

\maketitle

\begin{abstract}
Silicon double paddle osillators (DPO) have been successfully used for measuring the elastic properties of amorphous films down to 10 mK (see for example \cite{Fefferman16,Liu14}).  Until now, our group has used a wet dilution refrigerator for the lowest temperature measurements. We present measurements carried out on a Bluefors cryogen-free dilution refrigerator that demonstrate an extreme sensitivity of the thermal coupling of the DPO to its environment. These measurements show that it is necessary to enclose the DPO in a shield at the mixing chamber (MXC) temperature. Any gaps in the shield limit its effectiveness, even if there is no line-of-sight path to the DPO. In the absence of a cryogenic hermetic shield surrounding the DPO, turning off the pulse tube while maintaining the MXC and still temperatures leads to heating of the DPO. This demonstrates that any heating of the sample due to pulse tube vibrations is a less important effect.
\end{abstract}

\section{Introduction}
\label{intro}
Automated cryogen free dilution refrigerators provide a great deal of experimental space and generally reduce the effort required to carry out low temperature experiments, in part because the helium bath is replaced by a cryocooler, which is a pulse tube refrigerator (PT) in our case. One drawback of cryogen free dilution refrigerators is the higher level of vibrations due to the pulse tube \cite{Pelliccione13,Kalra16}. A sketch of our type of cryostat (BlueFors LD400) is shown in \cite{Kalra16}. The absence of a hermetic shield at 4 kelvin simplifies the design, but it can also complicate certain types of measurements, such as the one described below.

Silicon double paddle oscillators (DPO) are used to make highly sensitive measurements of the elastic properties of thin films \cite{Liu01,Spiel01,Fefferman10,Liu12}. We are using the DPO to make measurements of a-Si films. Initially, we were making measurements on a wet dilution refrigerator (now decommissioned) \cite{Fefferman16}. After transfering the setup to a cryogen-free dilution refrigerator, we observed a weaker temperature dependence of the DPO resonance than was observed on the wet refrigerator. In the following we demonstrate that the onset temperature of the levelling off can be reduced by adding a shield around the DPO at the mixing chamber temperature. However, even in the case where the shield has no visible gaps, the DPO remains sensitive to its environment, as demonstrated by the increase in its resonance frequency after stopping the pulse tube.

\section{Experimental Details}
\label{sec:1}
\begin{figure}
  \includegraphics[width=\textwidth]{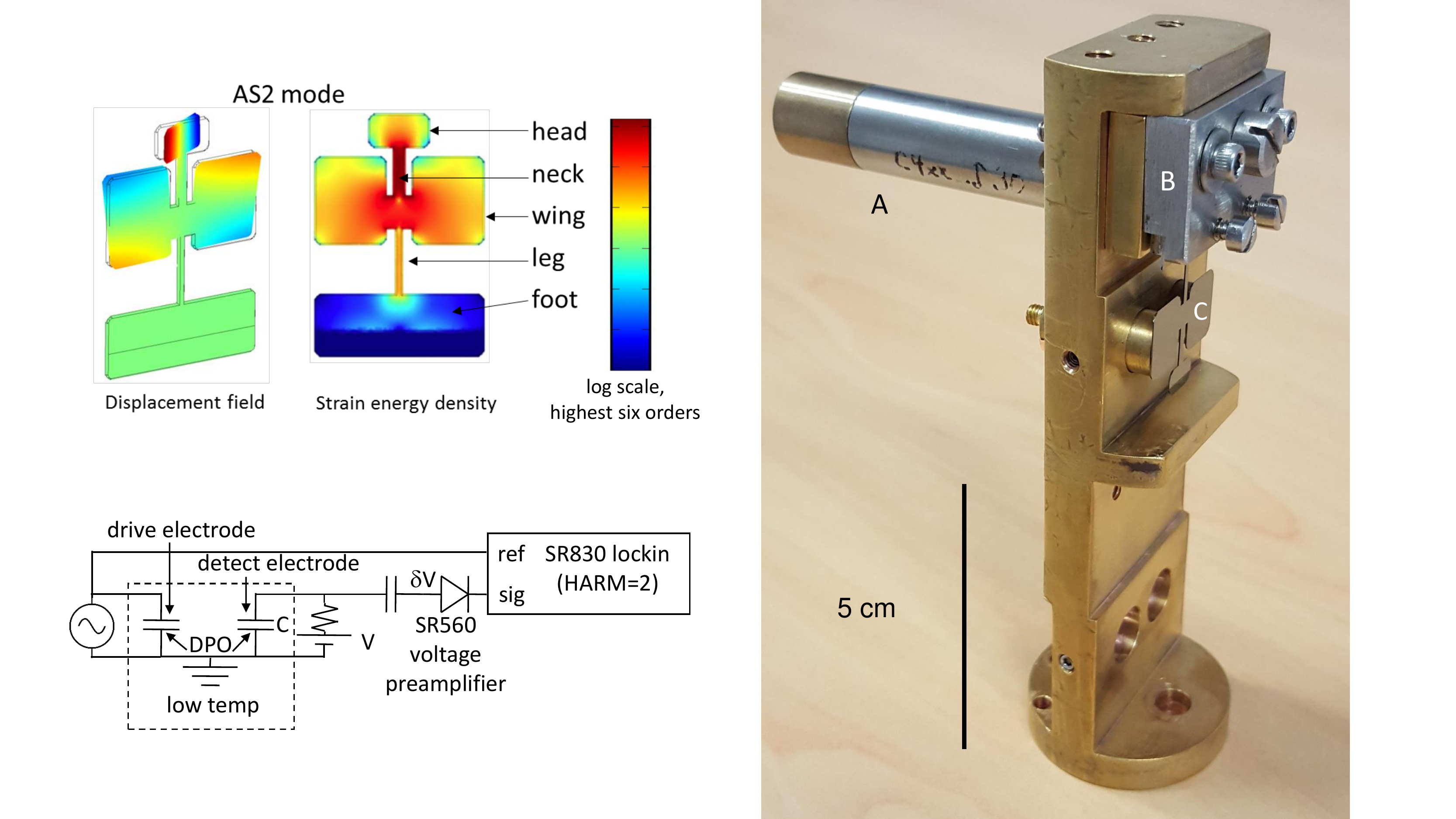}
\caption{Upper left: Calculation of the motion of a double paddle oscillator (DPO). A grounded platinum film deposited on the paddle is used for electrostatic drive and detection of its motion. It covers neither the neck, which is the sample region, nor the head. Also shown is the circuit used for drive and detection. Right: Photograph of the gold-plated copper electrode structure showing (A) Magnicon noise thermometer, (B) Invar part of the DPO clamping block, (C) DPO. The bottom of this structure is bolted to the MXC. (Color figure online.)}
\label{fig:1}
\end{figure}

The DPO has been characterized in detail in previous work \cite{Liu01,Spiel01,Fefferman10,Liu12}. It
is fabricated by wet etching a 300 $\mu$m thick wafer of crystalline silicon. Of the ten lowest resonant modes of the DPO, the AS2 mode around 5.5 kHz has the lowest
$Q^{-1}$, which reaches below 10$^{-8}$ at the lowest temeratures. The other
modes have $Q^{-1}$ $\approx$ 10$^{-6}$ at low temperatures. The exceptionally
low $Q^{-1}$ of the AS2 mode is due to the low strain amplitude at the
foot, where the DPO is clamped (Fig. \ref{fig:1}). In this mode, the strain is concentrated in the upper
torsion rod, called the neck, where the sample is deposited. A Au or Pt metal film is deposited on the wings, leg and foot of the DPO to
facilitate electrostatic drive and detection of the motion and to thermalize
the DPO. When the DPO is installed in its electrode structure, each wing is
separated from an electrode by a $\approx100$ $\mu$m gap, forming two of the
capacitors represented in the circuit (Fig. \ref{fig:1}). In our previous work \cite{Fefferman16},
the DPO was clamped in a large Invar block, whose low thermal expansion coefficient minimizes the risk
of cracking the DPO while cooling. The high thermal time constant of the Invar block also led to very long cooling times. Therefore we now use a half Invar/half copper block designed so that only the parts whose thermal contraction could squeeze the DPO were made of Invar. This block was in turn screwed to a copper electrode structure, which is screwed to the mixing chamber plate. In some cooldowns, a Magnicon noise thermometer was screwed to the electrode structure, as shown in Fig. \ref{fig:1}, to verify the absence of a significant thermal gradient across the electrode structure, even during the times when the pulse tube was temporarily turned off (see below).

The resonance frequency of the DPO was determined by the ringdown technique using the circuit shown in Fig. \ref{fig:1}.
First, we applied a drive voltage across the capacitor formed by the drive electrode and the
grounded metal film on the DPO. Since the force on the DPO is proportional to the square of the drive voltage, the frequency of the driving voltage was set to half the DPO resonant frequency. At the start of the ringdown,
the driving voltage was reduced by a factor of 250 and its frequency was increased by 10 mHz above the DPO resonance,
so that the AS2 mode was effectively undriven.
To detect the motion, the other electrode was biased to a voltage $V$. A blocking capacitor
protected the preamplifier from the large dc voltage. Due to a 10 M$\Omega$ resistor in series with the voltage source and the
100 M$\Omega$ input impedance of the preamplifier, the charge on the detection electrode was nearly constant over a period of the DPO oscillation. Thus the fluctuations $\delta V$ of the voltage across the detection capacitor
$C$ were
proportional to the capacitance fluctuations $\delta C$ caused by the motion of the
DPO: $\delta V/V=-\delta C/C$. The quantity $\delta V$ was measured with the ``Harm=2'' setting of an SR830 dual phase lockin amplifier. This setting causes
it to generate an internal reference signal at a frequency $f_{\mathrm{ref}}$ that is twice the
frequency of the signal output by the function generator. The lockin output the phase $\theta$ of $\delta V$
relative to the reference signal. The oscillation frequency of the DPO during the ringdown
(i.e. its resonant frequency $f_{\mathrm{res}}$) was determined using $f_{\mathrm{res}}=f_{\mathrm{ref}}+(1/2\pi)d\theta/dt$.

\section{Results and Discussion}
\begin{figure}
  \includegraphics[width=\textwidth]{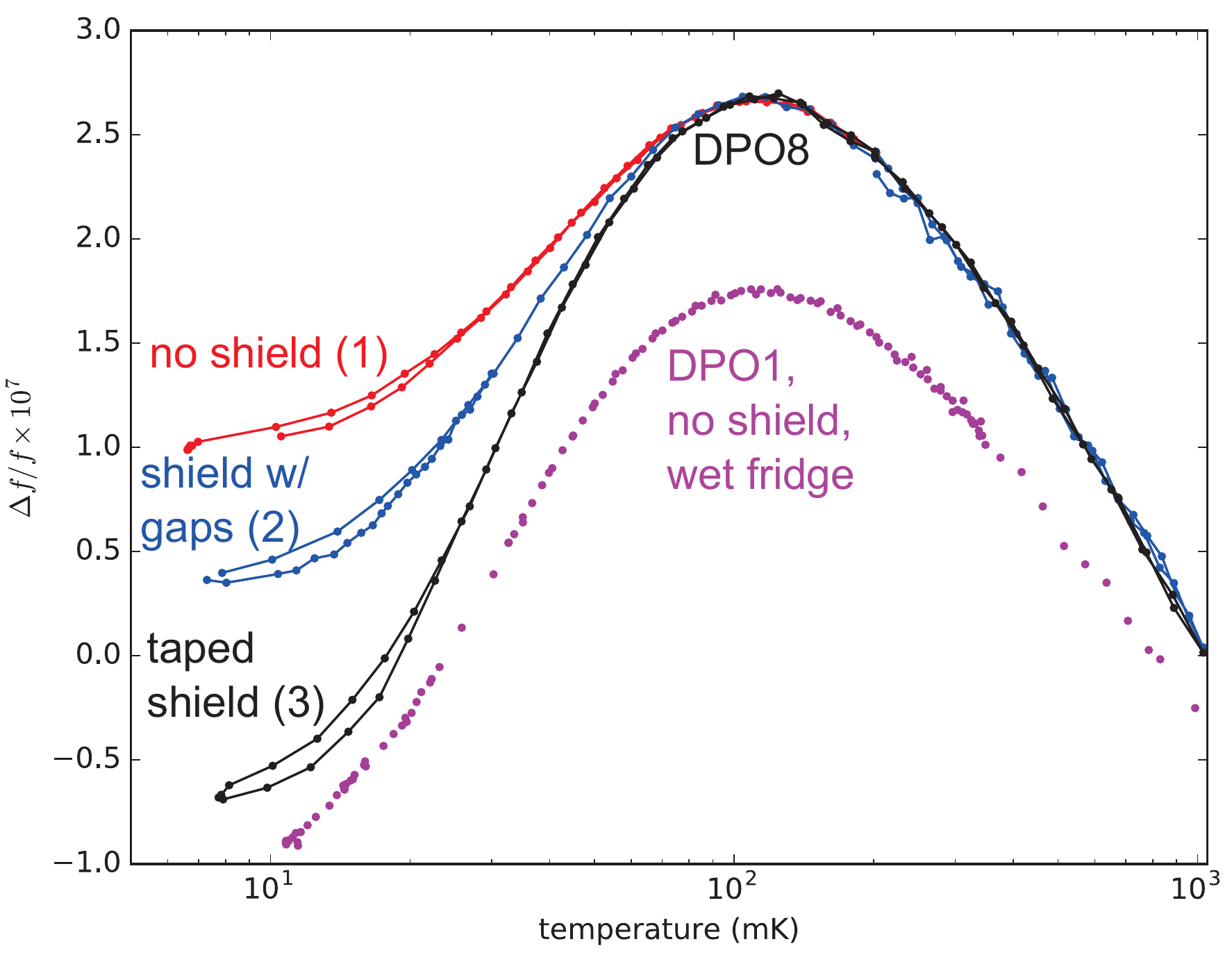}
\caption{Relative frequency shift of DPO1 and DPO8, each with a 300 nm thick amorphous Si film deposited at room temperature. The labels correspond to different shield configurations (see text). Warming and cooling curves are shown for the DPO8 measurements; the small hysteresis windows at the lowest temperatures are due to the thermal time constant of the Invar block or that of the DPO itself. (Color figure online.)}
\label{fig:2}       
\end{figure}

The amplitude of the response during the ringdown of a
bare DPO at 10 mK was shown in Fig. \ref{fig:1} of \cite{Fefferman16}. As
expected, it had a nearly perfect exponential time dependence. The corresponding $f_{\mathrm{res}}$ was also shown to be nearly constant during the ringdown. As in that case, all of the measurements presented here were made at sufficiently low strain to be in the linear regime. Figure \ref{fig:2} shows the temperature dependence of the resonant frequency
of the present DPO obtained from ringdown measurements like those shown in Fig.
\ref{fig:1} of \cite{Fefferman16}. This sample, called DPO8, has a 300 nm thick a-Si film deposited at room temperature. Figure \ref{fig:2} shows measurements of DPO8 in three configurations: (1) no mixing chamber shield surrounding the DPO, i.e., with a line of sight from the still shield to the DPO (2) a simple mixing chamber shield with a $\approx$ 1 cm gap but no line of sight from the still shield to the DPO, and (3) with all visible gaps in the simple shield closed with Cu tape. A temperature independent frequency offset was added to each of the DPO8 measurements so that the high temperature data overlap. Also shown is a measurement of another DPO, called DPO1, that also has a 300 nm thick a-Si film deposited at room temperature. This measurement was carried out in a wet frige with no shield at the mixing chamber, i.e., there was a line of sight from the DPO to the still shield.

It is clear that the thermalization of DPO8 greatly improved as its mixing chamber shield was improved. We now consider the possible sources of heat that are blocked by the shield. In the absence of a mixing chamber shield, we did not observe a significant change in the temperature dependence of a DPO resonance after the still temperature was changed from 630 to 900 mK. Therefore thermal radiation from the still is not significant and gas in the vacuum space is likely responsible for the decoupling in cases (1) and (2) and possibly in (3). We can now roughly estimate the heat required to produce the observed thermal decoupling. Since the sample region (the ``neck'') of the DPO is not coated with a metal film, the thermal gradient along the neck of the DPO depends on the thermal resistance of the crystalline Si substrate. This thermal resistance is uncertain because it depends on the unknown concentration of defects and the unknown roughness of the sample, which determines the probability of specular phonon scattering. We can only conclude that the phonon mean free path must be less than the length of the DPO neck, i.e, 6 mm. In the case where the head of the DPO is at 30 mK and the other end of the neck is at 10 mK, and the phonon mean free path is 6 mm, the heating of the head needed to sustain this gradient would be 0.2 nW \cite{Heron09}. The gas that is heating the DPO above the mixing chamber temperature could be residual atomospheric helium that was not removed when pumping the vacuum space at room temperature, atmospheric helium leaking in through the rubber seals of the vacuum can, or helium entering from small leaks in the dilution unit or pulse tube. Any other gases leaking into the cryostat should freeze when they collide with surfaces inside the cryostat, but the heat of absorption could cause additional vaporization in a cascade process that results in some heating of the DPO. However, this scenario is unlikely since we do not observe any change in the mass loading of the DPO over time. The sensitivity of the DPO resonance to mass loading is high: a helium layer with an effective thickness of 0.4 nm would cause a shift $df/f=4\times10^{-8}$.

When compared with the measurements of DPO8, the measurement of DPO1 in Fig. \ref{fig:2} shows that the heating of DPOs due to gas in the vacuum can is much worse in our cryogen-free cryostat than it was in the wet cryostat. This seems to be due to the absence of a hermetic seal at 4 kelvin in the cryogen-free cryostat. Indeed, the fact that a line of sight from the DPO to a surface with a temperature above that of the mixing chamber is not required for heating of the paddle by gas (see above) implies that the gas could be transporting heat from parts of the cryostat that are outside of the 4 kelvin shield. The levelling off of DPO1 below 20 mK is less than that of DPO8 even in the case with the taped shield, suggesting that we still need to improve the shield to reduce thermal decoupling to the level obtained on the wet fridge without a mixing chamber shield. The difference in the temperature dependence of the frequency of DPO1 and that of DPO8 with the taped shield above 20 mK is probably due to differences in the samples, not thermal decoupling.

\begin{figure}
  \includegraphics[width=\textwidth]{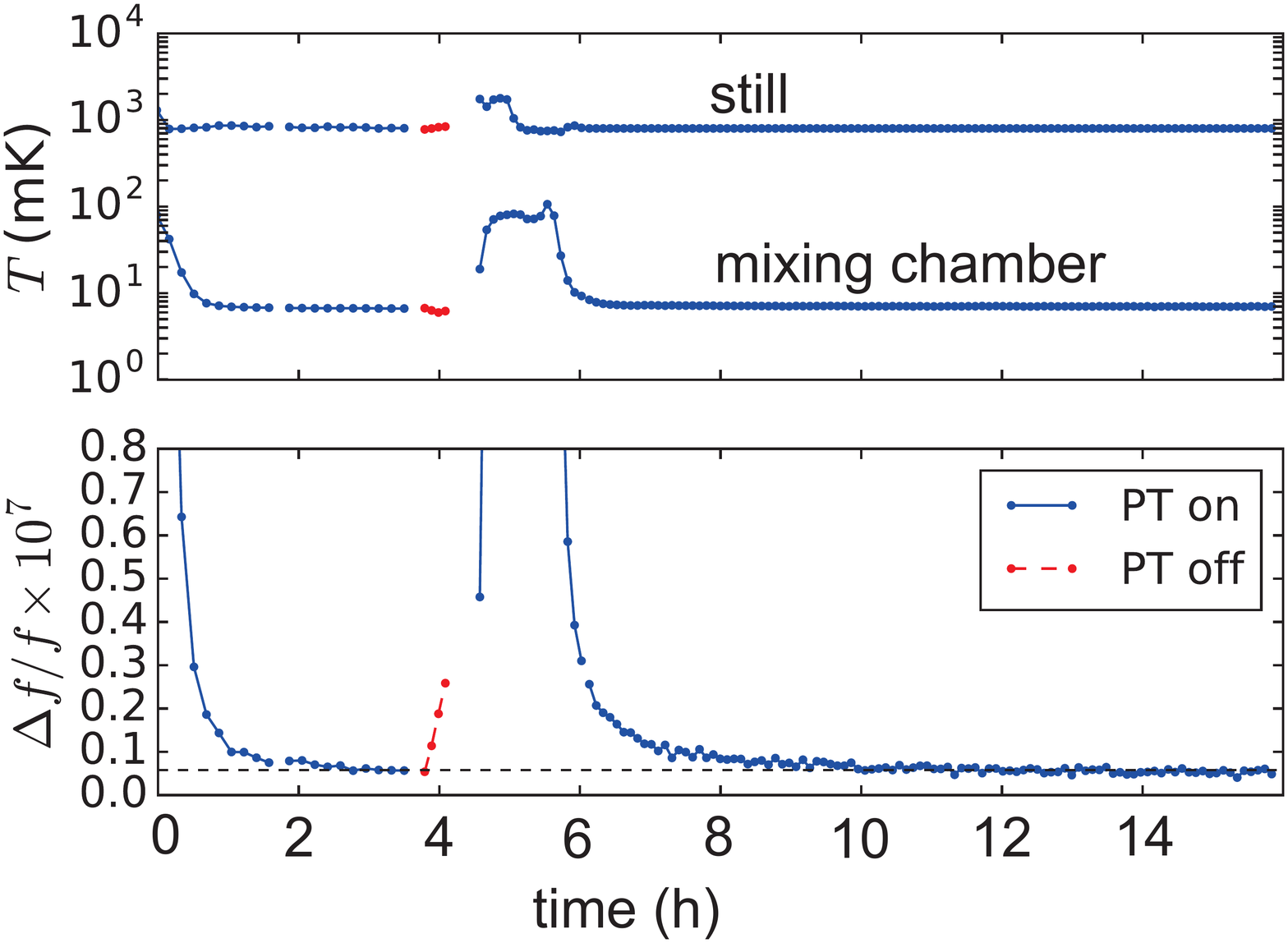}
\caption{Warming of the DPO upon stopping the pulse tube cooler (PT). The resonance frequency of the DPO
 is first allowed to relax to its minimum value with the cryostat at its base temperature. The PT is then stopped while the mixing chamber and still temperatures are temporarily maintained (red), leading to an increase in the DPO resonance frequency due to warming. Eventually the mixing chamber begins to warm. After restarting the PT and returning to the base temperature, the DPO resonance returns to its intial value (dashed black line is a horizontal guide to the eye). (Color figure online.)}
\label{fig:3}       
\end{figure}

Figure \ref{fig:3} shows the time dependence of still and mixing chamber temperatures along with that of the resonance frequency of DPO8 measured in shield configuration (3). The resonance frequency of the DPO
 is first allowed to relax to its minimum value with the cryostat at its base temperature. The PT is then stopped while the mixing chamber and still temperatures are temporarily maintained (red), leading to an increase in the DPO resonance frequency due to warming. Eventually the mixing chamber begins to warm. After restarting the PT and returning to the base temperature, the DPO resonance returns to its initial value. Apparently the warming that occurs after the PT is turned off is due to helium in the vacuum can that is transporting heat from surfaces that are normally at 4 or 50 K and which warm considerably after the PT is turned off. This helium warms the DPO despite many collisions with surfaces near the temperature of the mixing chamber as it gets under the Cu tape used to seal the DPO shield or in the space between screws and their threaded holes. That fact that the DPO returns to its original frequency at base temperature after the pulse tube is restarted shows that the gas causing the heating is helium, since other gases would freeze to the paddle and cause mass loading, to which the DPO is quite sensitive (see above).

These results imply that it may be necessary to use a cryogenic hermetic shield to reduce thermal decoupling of the DPO to the level obtained on the wet fridge. Alternatively, the taped shield (configuration 3) may be sufficient, and the levelling off below 20 mK present in this configuration could have another source such as pulse tube vibrations or the intrinsic temperature dependence of the shear modulus of the a-Si film. We are constructing a hermetic shield at the mixing chamber temperature to test these possibilities.

We acknowledge support from the ERC StG grant UNIGLASS No. 714692 and ERC CoG grant ULT-NEMS No. 647917 and from the US Office of Naval Research.

\bibliographystyle{spphys}       

\end{document}